\def\spacingset#1{\renewcommand{\baselinestretch}{#1}\small\normalsize}
\documentstyle[12pt]{article}
\begin{document}
\newcommand{\nc}{\newcommand}
\nc{\rnc}{\renewcommand}
\nc{\nt}{\newtheorem}
\nc{\be}{\begin}
\nc{\erf}[1]{$\ (\ref{#1}) $}
\nc{\rf}[1]{$\ \ref{#1} $}
\nc{\lb}[1]{\mbox {$\label{#1}$} }
\nc{\hr}{\hrulefill}
\nc{\noi}{\noindent}

\nc{\eq}{\begin{equation}}
\nc{\en}{\end{equation}}
\nc{\eqa}{\begin{eqnarray}}
\nc{\ena}{\end{eqnarray}}

\nc{\ra}{\rightarrow}
\nc{\la}{\leftarrow}
\nc{\da}{\downarrow}
\nc{\ua}{\uparrow}
\nc{\Ra}{\Rightarrow}
\nc{\La}{\Leftarrow}
\nc{\Da}{\Downarrow}
\nc{\Ua}{\Uparrow}

\nc{\uda}{\updownarrow}
\nc{\Uda}{\Updownarrow}
\nc{\lra}{\longrightarrow}
\nc{\lla}{\longleftarrow}
\nc{\llra}{\longleftrightarrow}
\nc{\Lra}{\Longrightarrow}
\nc{\Lla}{\Longleftarrow}
\nc{\Llra}{\Longleftrightarrow}

\nc{\mt}{\mapsto}
\nc{\lmt}{\longmapsto}
\nc{\lt}{\leadsto}
\nc{\hla}{\hookleftarrow}
\nc{\hra}{\hookrightarrow}
\nc{\lgl}{\langle}
\nc{\rgl}{\rangle}

\nc{\stla}{\stackrel{d}{\la}}
\nc{\pard}{\partial \da}
\nc{\gdot}{\circle*{0.5}}


\rnc{\baselinestretch}{1.2}      
\nc{\bl}{\vspace{1ex}}           
\rnc{\theequation}{\arabic{section}.\arabic{equation}}  
\newcounter{xs}
\newcounter{ys}
\newcounter{os}
\nt{thm}{Theorem}[section]
\nt{dfn}[thm]{Definition}
\nt{pro}[thm]{Proposition}
\nt{cor}[thm]{Corollary}
\nt{con}[thm]{Conjecture}
\nt{lem}[thm]{Lemma}
\nt{rem}[thm]{Remark}
\nc{\Poincare}{\mbox {Poincar$\acute{\rm e}$} }

\nc{\bA}{\mbox{${\bf A}$\ }}
\nc{\bB}{\mbox{${\bf B}$\ }}
\nc{\bC}{\mbox{${\bf C}$\ }}
\nc{\bD}{\mbox{${\bf D}$\ }}
\nc{\bE}{\mbox{${\bf E}$\ }}
\nc{\bF}{\mbox{${\bf F}$\ }}
\nc{\bG}{\mbox{${\bf G}$\ }}
\nc{\bH}{\mbox{${\bf H}$\ }}
\nc{\bI}{\mbox{${\bf I}$\ }}
\nc{\bJ}{\mbox{${\bf J}$\ }}
\nc{\bK}{\mbox{${\bf K}$\ }}
\nc{\bL}{\mbox{${\bf L}$\ }}
\nc{\bM}{\mbox{${\bf M}$\ }}
\nc{\bN}{\mbox{${\bf N}$\ }}
\nc{\bO}{\mbox{${\bf O}$\ }}
\nc{\bP}{\mbox{${\bf P}$\ }}
\nc{\bQ}{\mbox{${\bf Q}$\ }}
\nc{\bR}{\mbox{${\bf R}$\ }}
\nc{\bS}{\mbox{${\bf S}$\ }}
\nc{\bT}{\mbox{${\bf T}$\ }}
\nc{\bU}{\mbox{${\bf U}$\ }}
\nc{\bV}{\mbox{${\bf V}$\ }}
\nc{\bW}{\mbox{${\bf W}$\ }}
\nc{\bX}{\mbox{${\bf X}$\ }}
\nc{\bY}{\mbox{${\bf Y}$\ }}
\nc{\bZ}{\mbox{${\bf Z}$\ }}
\nc{\cA}{\mbox{${\cal A}$\ }}
\nc{\cB}{\mbox{${\cal B}$\ }}
\nc{\cC}{\mbox{${\cal C}$\ }}
\nc{\cD}{\mbox{${\cal D}$\ }}
\nc{\cE}{\mbox{${\cal E}$\ }}
\nc{\cF}{\mbox{${\cal F}$\ }}
\nc{\cG}{\mbox{${\cal G}$\ }}
\nc{\cH}{\mbox{${\cal H}$\ }}
\nc{\cI}{\mbox{${\cal I}$\ }}
\nc{\cJ}{\mbox{${\cal J}$\ }}
\nc{\cK}{\mbox{${\cal K}$\ }}
\nc{\cL}{\mbox{${\cal L}$\ }}
\nc{\cM}{\mbox{${\cal M}$\ }}
\nc{\cN}{\mbox{${\cal N}$\ }}
\nc{\cO}{\mbox{${\cal O}$\ }}
\nc{\cP}{\mbox{${\cal P}$\ }}
\nc{\cQ}{\mbox{${\cal Q}$\ }}
\nc{\cR}{\mbox{${\cal R}$\ }}
\nc{\cS}{\mbox{${\cal S}$\ }}
\nc{\cT}{\mbox{${\cal T}$\ }}
\nc{\cU}{\mbox{${\cal U}$\ }}
\nc{\cV}{\mbox{${\cal V}$\ }}
\nc{\cW}{\mbox{${\cal W}$\ }}
\nc{\cX}{\mbox{${\cal X}$\ }}
\nc{\cY}{\mbox{${\cal Y}$\ }}
\nc{\cZ}{\mbox{${\cal Z}$\ }}

\nc{\rightcross}{\searrow \hspace{-1 em} \nearrow}
\nc{\leftcross}{\swarrow \hspace{-1 em} \nwarrow}
\nc{\upcross}{\nearrow \hspace{-1 em} \nwarrow}
\nc{\downcross}{\searrow \hspace{-1 em} \swarrow}
\nc{\prop}{| \hspace{-.5 em} \times}
\nc{\wh}{\widehat}
\nc{\wt}{\widetilde}
\nc{\nonum}{\nonumber}
\nc{\nnb}{\nonumber}
 \nc{\half}{\mbox{$\frac{1}{2}$}}
\nc{\Cast}{\mbox{$C_{\frac{\infty}{2}+\ast}$}}
\nc{\Casth}{\mbox{$C_{\frac{\infty}{2}+\ast+\frac{1}{2}}$}}
\nc{\Casm}{\mbox{$C_{\frac{\infty}{2}-\ast}$}}
\nc{\Cr}{\mbox{$C_{\frac{\infty}{2}+r}$}}
\nc{\CN}{\mbox{$C_{\frac{\infty}{2}+N}$}}
\nc{\Cn}{\mbox{$C_{\frac{\infty}{2}+n}$}}
\nc{\Cmn}{\mbox{$C_{\frac{\infty}{2}-n}$}}
\nc{\Ci}{\mbox{$C_{\frac{\infty}{2}}$}}
\nc{\Hast}{\mbox{$H_{\frac{\infty}{2}+\ast}$}}
\nc{\Hasth}{\mbox{$H_{\frac{\infty}{2}+\ast+\frac{1}{2}}$}}
\nc{\Hasm}{\mbox{$H_{\frac{\infty}{2}-\ast}$}}
\nc{\Hr}{\mbox{$H_{\frac{\infty}{2}+r}$}}
\nc{\Hn}{\mbox{$H_{\frac{\infty}{2}+n}$}}
\nc{\Hmn}{\mbox{$H_{\frac{\infty}{2}-n}$}}
\nc{\HN}{\mbox{$H_{\frac{\infty}{2}+N}$}}
\nc{\HmN}{\mbox{$H_{\frac{\infty}{2}-N}$}}
\nc{\Hi}{\mbox{$H_{\frac{\infty}{2}}$}}
\nc{\Ogast}{\mbox{$\Omega_{\frac{\infty}{2}+\ast}$}}
\nc{\Ogi}{\mbox{$\Omega_{\frac{\infty}{2}}$}}
\nc{\Wedast}{\bigwedge_{\frac{\infty}{2}+\ast}}

\nc{\Fen}{\mbox{$F_{\xi,\eta}$}}
\nc{\Femn}{\mbox{$F_{\xi,-\eta}$}}
\nc{\Fp}{\mbox{$F_{0,p}$}}
\nc{\Fenp}{\mbox{$F_{\xi',\eta'}$}}
\nc{\Fuv}{\mbox{$F_{\mu,\nu}$}}
\nc{\Fuvp}{\mbox{$F_{\mu',\nu'}$}}
\nc{\cpq}{\mbox{$c_{p,q}$}}
\nc{\Drs}{\mbox{$\Delta_{r,s}$}}
\nc{\spq}{\mbox{$\sqrt{2pq}$}}
\nc{\Mcd}{\mbox{$M(c,\Delta)$}}
\nc{\Lcd}{\mbox{$L(c,\Delta)$}}
\nc{\Wxv}{\mbox{$W_{\chi,\nu}$}}
\nc{\vxv}{\mbox{$v_{\chi,\nu}$}}
\nc{\dd}{\mbox{$\widetilde{D}$}}

\nc{\diff}{\mbox{$\frac{d}{dz}$}}
\nc{\Lder}{\mbox{$L_{-1}$}}
\nc{\bone}{\mbox{${\bf 1}$}}
\nc{\px}{\mbox{${\partial_x}$}}
\nc{\py}{\mbox{${\partial_y}$}}
\setcounter{equation}{0}

\vspace{1in}
\begin{center}
{{\LARGE\bf Algebraic and Geometric Structures in String Backgrounds }}
\end{center}
\addtocounter{footnote}{0}
\vspace{1ex}
\begin{center}
Bong H. Lian\\
Department of Mathematics\\
Harvard University\\
Cambridge, MA 02138\\
lian$@$math.harvard.edu
\end{center}
\begin{center}
Gregg J. Zuckerman\\
Department of Mathematics\\
Yale University\\
New Haven, CT 06520\\
gregg$@$math.yale.edu
\end{center}
\addtocounter{footnote}{0}
\footnotetext{B.H.L. is supported by grant DE-FG02-88-ER-25065.
G.J.Z. is supported by NSF Grant DMS-9307086 and DOE Grant
DE-FG02-92-ER-25121.}

\vspace{1ex}
\begin{quote}
{\footnotesize
ABSTRACT.  We give a brief introduction to the study of the algebraic
structures -- and their geometrical interpretations -- which arise in the BRST
construction of a conformal string background. Starting from the chiral
algebra $\cA$ of a string background, we consider a number of
elementary but universal operations on the chiral algebra.
{}From these operations we deduce a certain
fundamental odd Poisson structure, known as a Gerstenhaber algebra, on the BRST
cohomology of $\cA$. For the 2D string background,
the correponding G-algebra can be partially described in term of a geometrical
G-algebra of the affine plane $\bC^2$. This paper will appear in the
proceedings of {\it Strings 95}.
}
\end{quote}

\addtocounter{footnote}{0}

\section{Introduction}

The BRST formalism is a widely, if not universally, recognized approach
to the imposition of the Virasoro constraints in string theory
(for some early works, see \cite{KO}\cite{FMS}\cite{W1}\cite{Fe}\cite{FGZ}).
Over the last dozen years physicists and mathematicians alike have
pondered the BRST-structure of string backgrounds, both abstract and
concrete. (For further discussion of the BRST formalism in string and string
field theory, see the paper in this volume by Zwiebach.) During the
same period, conformal field theory techniques have played
an ever increasing role in the theory of string backgrounds (see
\cite{BPZ}\cite{MS}.) For a long time, the general
theory has been dominated by the standard construction of
ghost number one BRST ivariant fields from dimension one
primary matter fields  (see for example \cite{FMS}).
It is well known that such standard
invariant fields form a Lie algebra,
at least modulo exact fields.

A number of research groups have understood that the operator
product expansion for the background chiral algebra leads to a
much richer algebraic structure in the {\it full} BRST cohomology,
where all ghost numbers are on an equal footing
\cite{Wi3}\cite{WZ}\cite{LZ9}\cite{SP}\cite{Hu}.
 In \cite{LZ9}, the
authors have recognized that the full BRST cohomology of a
conformal string background has the structure of
a Batalin-Vilkovisky algebra, which is a special type of a
Gerstenhaber algebra, or G-algebra. A G-algebra is a
generalization of a Poisson algebra, which incorporates
simultaneously the structure of a commutative algebra
and a Lie algebra \cite{Gers1}\cite{Gers2}\cite{GGS}.
(For a further discussion of BV- and G-algebras in
the context of W-string backgrounds, see the paper in this
volume by McCarthy et al.)

The particular G-algebra of the 2D string background is especially
complex and has probably made no previous appearance in pure
mathematics. In particular, in ghost number one BRST cohomology
we obtain a (noncentral) extension of the Lie algebra of vector fields
in the plane by an infinite dimensional abelian Lie algebra \cite{LZ9}. In
this sense, string theory has enriched our understanding of
algebraic structures. At the same time, there is a connection
between the G-algebra for 2D strings and the anti-bracket
formalism. The latter appears in both the work of Schouten-Nijenhuis
\cite{nijenhuis}
on the tensor calculus as well as the work of Batalin-Vilkovisky \cite{bv}
\cite{Sch}
on the quantization of constrained field theories. Thus,
string theory has also illuminated our understanding of
geometrical structures.

As a biproduct of our structural analysis of the 2D string background,
we obtain a new and very simple description of an explicit basis for
the (chiral) BRST cohomology. This basis appears already in the work of
Witten and Zwiebach \cite{WZ}. Our enumeration of the basis exploits
the full strength of the G-algebra structure, as well as a beautiful
representation of the group $SL(2,\bC)$ in the cohomology.
 We predict that the fruitful interaction between mathematics and
string theory will continue to be as exciting in the future as it
has been over the last twenty-six years.

{\it A note to our mathematical friends:}  In section three below we
use the standard contour integral formalism of conformal field
theory. This can be understood algebraically as follows: assume
that the chiral algebra $\cA$ is actually a
commutative quantum operator algebra in the sense of \cite{LZ14}.
There are an infinite number of bilinear operations in $\cA$ -- the
circle products. The relationship between the circle products and
contour integrals is very simple:
\eq
\oint_{C_w}\cO_1(z)\cO_2(w)(z-w)^ndz=\cO_1(w)\circ_n\cO_2(w).
\en
With this translation, it is now possible to read the current
paper as a follow-up to our paper \cite{LZ14} on CQOAs. For the
closely related point of view of vertex operator algebras,
see \cite{Bor}\cite{FLM}.

In this paper, we begin with a brief review of the BRST contruction
of conformal string backgrounds. The 2D string background will be
a fundamental example throughout our discussion. Using the
elementary notions such as the normal ordered product and the
descent operation, we show how to construct some fundamental
algebraic structures on the BRST cohomology. In the case of
the 2D string background, these structures can be described
to a large extent in terms of the geometric G-algebra of the
affine plane $\bC^2$. We also describe an explicit basis for
the BRST cohomology using the $SL(2,\bC)$ group action on
$\bC^2$.

\section{BRST Formalism}

The chiral algebra of a conformal string background is of the form
\eq
\cA=\cA^{ghost}\otimes\cA^{matter}
\en
where $\cA^{ghost}$, $\cA^{matter}$ are respectively the chiral algebras
of the ghost and the matter sectors. Thus explicitly an element $\cO$ of $\cA$
is a finite
sum of holomorphic fields of the form $P(b,\partial b,\cdots,c,\partial
c,\cdots) \Phi$, where $P$ is a normal ordered differential
polynomial of the ghost fields
$b,c$; $\Phi$ is a field in the matter sector. The respective (holomorphic
part of the) stress energy tensors of the two sectors are
$L^{ghost},L^{matter}$
whose central charges are respectively $c=26$, and $c=-26$. The BRST current is
\eq
J=c(L^{matter}+\half L^{ghost}),
\en
and the ghost number current is
\eq
F=cb.
\en

Let's recall a few basics of the BRST formalism. The BRST charge
\eq
Q=\oint_{C_0}J(z)dz
\en
has the property that $[Q,Q]=2Q^2=0$, where square brackets $[\cdot,\cdot]$
denotes the super commutator. Here $C_0$ is a small contour around $0$.
A BRST invariant field $\cO$ is one which satisfies
$[Q,\cO]=0$; and a BRST exact field is one which is
 of the form $\cO=[Q,\cO']$ for some
field $\cO'$. The BRST cohomology of the string background $\cA$ is
the quotient space
\eq
H^*(\cA)=\mbox{$\{Q$-invariant fields$\}/\{Q$-exact fields$\}$}
\en
which is graded by the ghost number $*$.

A standard way
to obtain BRST invariants is as follows.
Let $V$ be a primary field of dimension one in $\cA^{matter}$. Then
both $cV$ and $c\partial cV$ are BRST invariants of ghost number 1,2
respectively. The fields $\cO=1$ and $\cO=c\partial c\partial^2 c$ are two
universal BRST invariants known from the bosonic string theory.

\subsection{2D String Background}

The chiral algebra $\cA_{2D}$ of the 2D string background is generated
by the fields $b,c,\partial X,\partial\phi$, $e^{\pm(\pm iX-\phi)/\sqrt{2}}$,
where $X,\phi$ are free bosons with the usual OPEs. Their corresponding
stress energy tensors are:
\eqa
L^X&=&-\half(\partial X)^2\nnb\\
L^\phi&=&-\half(\partial\phi)^2+\sqrt{2}\partial^2\phi
\ena
whose respective central charges are $c=1, c=25$.

We now describe some BRST invariants of the  chiral algebra $\cA_{2D}$.
Since $e^{(\pm iX+\phi)/\sqrt{2}}$ are matter primary fields of dimension one,
we immediately obtain some standard BRST invariants $-ce^{(\pm
iX+\phi)/\sqrt{2}}$. We denote them by $Y^+_{1/2,\pm1/2}$. There are
also {\it exotic}
BRST invariants which cannot be obtained in the above standard way. For
example, in ghost number zero we have
\eq
\cO_{1/2,\pm1/2}=\left(cb+\frac{i}{\sqrt{2}}(\pm\partial
X-i\partial\phi)\right)
e^{(\pm iX-\phi)/\sqrt{2}}.
\en
These BRST invariants were identified in \cite{LZ4} (see also \cite{BMP}),
and their explicit formulas were given in \cite{Wi3}\cite{WZ}. Explicit
formulas for infinitely many BRST invariants of the 2D string background
are also known. See \cite{BMP}\cite{MMS}\cite{WuZhu}\cite{WZ}.

\section{Fundamental Operations}

Using the antighost field $b$ and contour integral, one can define a number
of useful operations on the fields $\cO$ in a string background. Given
a dimension $h$ field $\cO$,
we attach to it a field of dimension $h+1$
\eq
\cO^{(1)}(w)=\oint_{C_w}b(z)\cO(w)dz
\en
where $C_w$ is a small coutour around $w$.
We call this linear operation, which reduces ghost number by one, the descent
operation. Similarly, to $\cO$ we can attach a field
of dimension $h$
\eq
\Delta \cO(w)=\oint_{C_w}b(z)\cO(w)(z-w)dz.
\en
This linear operation is called the Delta operation.

There are two important bilinear operations defined as follows. The first
one is the dot product (a.k.a. the normal ordered product). Given two fields
$\cO_1,\cO_2$, we define
\eq
\cO_1(w)\cdot\cO_2(w)=\oint_{C_w}\cO_1(z)\cO_2(w)(z-w)^{-1}dz.
\en
Under the dot product, both the conformal dimension and the ghost number are
additive.

We define also the bracket operation
\eq
\{\cO_1(w),\cO_2(w)\}=\oint_{C_w}\cO_1^{(1)}(z)\cO_2(w)dz.
\en
Note that conformal dimension is additive, while ghost number is
shifted by -1 under the bracket operation. This operation was implicit
in \cite{Wi3}\cite{WZ}, and was studied in general in \cite{LZ9}.

Let
\eqa
Q*\cO(w)&=&\oint_{C(w)}J(z)\cO(w)dz=[Q,\cO(w)]\nnb\\
\Sigma*\cO(w)&=&\oint_{C(w)}L^{total}(z)\cO(w)(z-w)dz.
\ena
Then we have the following algebra of operations:
\eqa\lb{3.13}
\partial\cO&=&{[Q,\cO^{(1)}]}\nnb\\
\Delta^2&=&0\nnb\\
(Q*)^2&=&0\nnb\\
{[Q*,\Delta]}&=&\Sigma*\nnb\\
\Sigma*\cO&=&n\cO \mbox{ iff  the conformal dimension of $\cO$ is $n$.}
\ena
Moreover, we have the following identities \cite{LZ9}:
\eqa\lb{3.14}
&(a)& Q*(\cO_1\cdot\cO_2)=(Q*\cO_1)\cdot\cO_2+(-1)^{|\cO_1|}\cO_1\cdot
(Q*\cO_2)\nnb\\
&(b)& Q*\{\cO_1,\cO_2\}= \{Q*\cO_1,\cO_2\}+(-1)^{|\cO_1|-1}
\{\cO_1,Q*\cO_2\}\nnb\\
&(c)& (-1)^{|\cO_1|}\{\cO_1,\cO_2\}=
\Delta(\cO_1\cdot\cO_2)-(\Delta\cO_1)\cdot\cO_2
-(-1)^{|\cO_1|}\cO_1\cdot\Delta\cO_2\nnb\\
\ena
where $|\cO|$ denotes the ghost number of $\cO$.
Note that the first equation in \erf{3.13} is known as the
descent equation \cite{WZ}.

\subsection{Induced algebraic structures in BRST cohomology}

Let $[\cO], [\cO_1], [\cO_2]$ be cohomology classes in $H^*(\cA)$.
By virtue of the identities \erf{3.13} and \erf{3.14}, the above
operations induce the following well-defined operations on cohomology:
\eqa
\Delta [\cO]&=&[\Delta\cO_{[0]}]\nnb\\
{[\cO_1]}\cdot{[\cO_2]}&=&{[\cO_1\cdot\cO_2]}\nnb\\
\{[\cO_1],[\cO_2]\}&=& {[\{\cO_1,\cO_2\}]}.
\ena
Here $\cO_{[0]}$ is the projection of $\cO$ onto the subspace of $\cA$
consisting of fields of zero conformal dimension.
For example, given the standard BRST cohomology classes $\cO_i=cV_i$, $i=1,2$,
where the $V_i$ are matter primary fields of conformal dimension one, we have
\eq
\{[\cO_1],[\cO_2]\}=[\cO_3]
\en
where
\eq
\cO_3(w)=c(w)\oint_{C_w}V_1(z)V_2(w)dz.
\en

\subsection{Chiral ground ring}

Since ghost number is additive under dot product in cohomology, it follows
that $H^0(\cA)$ is closed under this product.

\be{thm}
The dot product in $H^0(\cA)$ is commutative and associative.
\end{thm}
The commutative associative algebra $H^0(\cA)$ is called the chiral
ground ring of $\cA$.
For example it is shown in \cite{Wi3}\cite{LZ9} that in the case
$\cA=\cA_{2D}$, the ground ring is the polynomial algebra generated by
the classes ${[\cO_{1/2,1/2}]}$ and ${[\cO_{1/2,-1/2}]}$.

\section{Fundamental Identities}

\be{thm}\cite{LZ9}
Let $u,v,t$ be classes in $H^*(\cA)$ of ghost numbers $|u|,|v|,|t|$
respectively. Then we have\\
(a) $u\cdot v=(-1)^{|u||v|}v\cdot u$\\
(b) $(u\cdot v)\cdot t=u\cdot(v\cdot t)$\\
(c) $\{u,v\}=-(-1)^{(|u|-1)(|v|-1)}\{v,u\}$\\
(d)
$(-1)^{(|u|-1)(|t|-1)}\{u,\{v,t\}\}
+(-1)^{(|t|-1)(|v|-1)}\{t,\{u,v\}\}
+(-1)^{(|v|-1)(|u|-1)}\{v,\{t,u\}\}=0$\\
(e) $\{u,v\cdot t\}=\{u,v\}\cdot t +(-1)^{(|u|-1)|v|}v\cdot\{u,t\}$\\
\end{thm}

\be{dfn}
A G-algebra is an integrally graded vector space $A^*$ with dot
and bracket products satisfying
$|u\cdot v|=|u|+|v|$, $|\{u,v\}|=|u|+|v|-1$ and the five identities
(a) through (e).
\end{dfn}
For references on the mathematical theory of G-algebras, see
\cite{Gers1}\cite{Gers2}\cite{GGS}\cite{ks}.

\be{pro}
The BRST cohomology $H^*(\cA)$ of a string background $\cA$ satisfies
further fundamental identities:\\
(f) $(-1)^{|u|}\{u,v\}=\Delta(u\cdot v)-(\Delta u)\cdot v-(-1)^{|u|}u\cdot
\Delta v$\\
(g) $|\Delta|=-1$ and $\Delta^2=0$.
\end{pro}
\be{dfn}
A BV algebra is a G-algebra, $A^*$, equipped with a linear operation $\Delta$
satisfying (f) and (g).
\end{dfn}
For references on the mathematical theory of BV algebras, see
\cite{Koszul}\cite{ks}\cite{Sch}\cite{W2}.
For related papers on algebraic structure in BRST cohomology, see
\cite{FGZ}\cite{GJ}\cite{Hu}\cite{KSV}\cite{KSV2}\cite{LZ10}
\cite{M}\cite{Pen}\cite{SP}\cite{WuZhu}.

\section{G-algebra of the 2D String Background}\lb{galgebra}

An important example of a G-algebra is the one given by the BRST cohomology of
the 2D string background:
\be{thm}
As a G-algebra, $H^*(\cA_{2D})$ is generated by the four classes
${[\cO_{1/2,\pm 1/2}]}$, ${[Y^+_{1/2,\pm 1/2}]}$. Moreover, $H^p(\cA_{2D})$
vanishes except for $p=0,1,2,3$.
\end{thm}
Note that $\{ {[Y^+_{1/2,- 1/2}]}, {[Y^+_{1/2, 1/2}]} \}=
{[ce^{\sqrt{2}\phi}]}\neq0$. (See \cite{LZ9}.)

We now describe  the structure of the G-algebra $H^*(\cA_{2D})$ in terms of a
well-known geometrical G-algebra. Let $M$ be a
smooth manifold. Then the space $V^*(M)$ of polyvector fields
(ie. antisymmetric contravariant tensor fields) on
$M$ admits the structure of a G-algebra, known as the Schouten algebra of $M$.
 The dot product of a $p$-vector field $P$ and a $q$-vector $Q$ is given by
\eq
(P\cdot Q)^{\nu_1\cdots\nu_p\mu_1\cdots\mu_q}
=P^{[\nu_1\cdots\nu_p} Q^{\mu_1\cdots\mu_q]}.
\en

Now let $P,Q$ be elements of $V^{p+1}(M), V^{q+1}(M)$ respectively.
The Schouten bracket can be described as follows: in local coordinates the
Schouten bracket $[P,Q]_S$ is given by \cite{nijenhuis}
\eq
[P,Q]_S^{\nu_1\cdots\nu_p\lambda\mu_1\cdots\mu_q}=
(p+1)P^{\rho [\nu_1\cdots\nu_p}\partial_\rho Q^{\lambda\mu_1\cdots\mu_q] }
-(q+1)Q^{\rho [\mu_1\cdots\mu_q}\partial_\rho P^{\lambda\nu_1\cdots\nu_p] }.
\en
Note that if $p=q=0$, then $[P,Q]_S$ is the ordinary Lie bracket of the vector
fields $P,Q$.

Let $A^*(\bC^2)$ be the holomorphic polyvector fields on $\bC^2$ with
polynomial coefficients. As a linear space, $A^*(\bC^2)$ is the super
polynomial algebra $\bC[x,y,\partial_x,\partial_y]$, where $x,y$ have ghost
number zero and
$\partial_x,\partial_y$ have ghost number one.

\be{thm}
The assignment ${[\cO_{1/2, 1/2}]}\mapsto x$,  ${[\cO_{1/2, -1/2}]}\mapsto y$,
${[Y^+_{1/2, -1/2}]}\mapsto \partial_x$, ${[Y^+_{1/2, 1/2}]}\mapsto \partial_y$
extends to a G-algebra
homomorphism $\psi$ of $H^*(\cA_{2D})$ onto $A^*(\bC^2)$.
\end{thm}
For details on this result, see \cite{LZ9}.

Introduce the notations $x^*=\partial_x$, $y^*=\partial_y$. Define the linear
operation on $A^*(\bC^2)$: $D=\frac{\partial}{\partial x}
\frac{\partial}{\partial x^*} +\frac{\partial}{\partial y}
\frac{\partial}{\partial y^*}$.
\be{pro}
(a) $A^*(\bC^2)$ equipped with $D$ is a BV algebra.\\
(b) For $u$ in $H^*(\cA_{2D})$, $\psi(\Delta u)=- D(\psi u)$.
\end{pro}
(See \cite{LZ9}\cite{WZ}.)

Introduce the notations $\tilde{x}={[\cO_{1/2, 1/2}]}$,  $\tilde{y}={[\cO_{1/2,
-1/2}]}$,
$\tilde{\partial_x}={[Y^+_{1/2, -1/2}]}$, $\tilde{\partial_y}={[Y^+_{1/2,
1/2}]}$,
 $\tilde{J}_+=\tilde{x}\tilde{\partial_y}$,
$\tilde{J}_0=\tilde{x}\tilde{\partial_x}-\tilde{y}\tilde{\partial_y}$,
$\tilde{J}_-=\tilde{y}\tilde{\partial_x}$. Similarly let
$J_+,J_0,J_-\in A^*(\bC^2)$ be
defined by analogous formulas but without the tildes.
\be{pro}
$Span\{\tilde{J}_+, \tilde{J}_0, \tilde{J}_-\}$,
$Span\{J_+,  J_0, J_-\}$ are both closed under the bracket $\{\cdot,\cdot\}$
and are isomorphic to the Lie
algebra $sl(2,\bC)$.
\end{pro}
(See \cite{Wi3}\cite{WZ}\cite{LZ9}\cite{BMP}.)

Now introduce an $sl(2,\bC)$-action on $H^*(\cA_{2D})$ by
$\tilde{J}_a*\cO=\{\tilde{J}_a, \cO\}$ where $a=+,0,-$, $\cO\in H^*(\cA_{2D})$.
Similarly, introduce an $sl(2,\bC)$-action on $A^*(\bC^2)$
 by $J_a*X=\{J_a, X\}$, $X\in A^*(\bC^2)$.
We make the observation (see \cite{LZ9}\cite{BMP}) that
$\psi:H^*(\cA_{2D})\ra A^*(\bC^2)$ intertwines actions of $sl(2,\bC)$.

\be{pro} \cite{LZ9}
$ker\ \psi$ is a $G$-ideal with vanishing dot product and bracket product.
\end{pro}
That is, for $\cO\in H^*(\cA_{2D})$, $\cO',\cO''\in ker\ \psi$,
we have $\cO\cdot\cO', \{\cO,\cO'\}\in ker\ \psi$, and
$\cO'\cdot\cO''=\{\cO',\cO''\}=0$.

\be{pro} \cite{LZ9}
$ker\ \psi$ as a $G$-ideal is generated by the class ${[ce^{\sqrt{2}\phi}]}$.
\end{pro}
This means that ${[ce^{\sqrt{2}\phi}]}\in ker\ \psi$, and the smallest subspace
containing ${[ce^{\sqrt{2}\phi}]}$ and
stable under the action of $H^*(\cA_{2D})$ by both the
dot product and the bracket product is the whole $ker\ \psi$ itself.

Let's give an explicit basis for $ker\ \psi$. Introduce the following linear
operations on cohomology classes $[\cO]$:
\eqa
A[\cO]&=&\{\tilde{\partial_x},[\cO]\}\nnb\\
B[\cO]&=&\{\tilde{\partial_y},[\cO]\}\nnb\\
C[\cO]&=&\tilde{\partial_x}\cdot[\cO]\nnb\\
D[\cO]&=&\tilde{\partial_y}\cdot[\cO].
\ena
Fix $\cK=ce^{\sqrt{2}\phi}$. Then we have
\be{pro}
$ker\ \psi$ as a vector space has a basis consisting of the following classes:
\eqa
&ghost\ no.\ 1:& A^{s-n}B^{s+n}[\cK]\nnb\\
&ghost\ no.\ 2:& (s-n)A^{s-n-1}B^{s+n}C[\cK]+
(s+n)A^{s-n}B^{s+n-1}D[\cK]\nnb\\
&ghost\ no.\ 2:& A^{s-n+1}B^{s+n}D[\cK]-
A^{s-n}B^{s+n+1}C[\cK]\nnb\\
&ghost\ no.\ 3:& A^{s-n}B^{s+n}CD[\cK]
\ena
where $s=0,\half,1,...,$ and $n=-s,-s+1,..,s$.
\end{pro}
We observe that the above basis vectors are in fact weight vectors for
the $sl(2,\bC)$ action which we have previously described. Here $s,n$
are respectively the total spin and axial spin quantum numbers of the
$sl(2,\bC)$ representation.

For completeness, we list here a basis for the algebra
$A^*(\bC^2)\cong H^*(\cA_{2D})/ker\ \psi$:
\be{eqnarray}
&ghost\ no.\ 0:&x^{s-n}\cdot y^{s+n}\nnb\\
&ghost\ no.\ 1:&\px(x^{s-n}\cdot y^{s+n})\cdot\py
-\py(x^{s-n}\cdot y^{s+n})\cdot\px\nnb\\
&ghost\ no.\ 1:&x^{s-n}\cdot y^{s+n}\cdot(x\cdot\px+y\cdot\py)\nnb\\
&ghost\ no.\ 2:&x^{s-n}\cdot y^{s+n}\cdot\px\cdot\py
\end{eqnarray}
where $s=0,\half,1,...,$ and $n=-s,-s+1,..,s$.

\section{Conclusion}

Already in \cite{Wi3}, Witten begins to give a geometrical interpretation of
the BRST cohomology of the 2D string background.  He shows that the ghost
number zero chiral BRST cohomology is isomorphic to the algebra of holomorphic
polynomials on complex 2-space (see section \ref{galgebra}).  Witten and
Zwiebach note that under the dot product, the BRST classes $\tilde{x}$,
$\tilde{y}$,
$\tilde{\partial_x}$, $\tilde{\partial_y}$ generate a graded commutative
associative algebra $H(+)$ isomorphic to $A^*(\bC^2)$, which is interpreted
geometrically as holomorphic polynomial polyvector fields on complex 2-space.
However, since $H(+)$ is clearly not closed under
the bracket operation, it is more difficult to give a purely geometrical
meaning to that operation.  It is also
tricky to give geometical meaning to $ker\ \psi$,
 which constitutes the other "half" of the BRST cohomology.

The point of view taken in our earlier paper \cite{LZ9} as well as in this
lecture is that the homomorphism $\psi$ provides us with a geometrical
interpretation of the quotient G-algebra, $H^*(\cA_{2D})/ ker\ \psi$.  For
example, the $sl(2, \bC)$ action
on the quotient can now be thought of as the infinitesimal counterpart to the
geometrical action of the Lie group
$SL(2, \bC)$ on $A^*(\bC^2)$.  It is interesting already to consider the orbit
structure of $SL(2, \bC)$ on $\bC^2$:  there are two
orbits, neither of which is free:  the first consists of the origin only, and
the second consists of all nonzero
points.  Observe that $\bC^2-\{0\}=SL(2,\bC)/N$ where $N$ is the subgroup of
upper triangular unipotent matrices.

For any complex semisimple Lie group $G$, we can consider an analog of the open
orbit of $SL(2, \bC)$ in $\bC^2$:  let the $N$ be a maximal unipotent subgroup
of $G$; that is, think of $G$ as a group of complex matrices (we can always do
this), and let $N$ be maximal among subroups of $G$ that consist of matrices of
the form $I + a\ nilpotent\ matrix$.  The coset space $G/N$ is called in
mathematics the base affine space of $G$.  In the lecture by McCarthy you will
see a far reaching generalization of our 2D string background in which a simple
group $G$ of ADE type replaces the $SL(2, \bC)$ that appears in the theory of
the 2D string background.

{\bf Acknowledgements.} The second author would like to thank the organizers'
invitation to lecture at the {\it Strings 95} conference. He would also like
to thank G. Moore and J. McCarthy for their helpful suggestions.

\end{document}